# Flow Topology Optimization at High Reynolds Numbers Based on Modified Turbulence Models

Chenyu Wu[1], Yufei Zhang[2]

(School of Aerospace Engineering, Tsinghua University, Beijing, China, 100084)

**Abstract:** Flow topology optimization (ToOpt) based on Darcy's source term is widely used in the field of ToOpt. It has a high degree of freedom, making it suitable for conceptual aerodynamic design. Two problems of ToOpt are addressed in this paper to apply the ToOpt method to high-Reynolds-number turbulent flow that is often encountered in aerodynamic design. First, a strategy for setting Darcy's source term is proposed based on the relationship between the magnitude of the source term and some characteristic-variables of the flow (length scale, freestream velocity and fluid viscosity). Second, two modified turbulence models including the modified Launder-Sharma $k-\epsilon$ (LSKE) model and the modified shear-stress-transport (SST) model that consider the influence of Darcy's source term on turbulence are developed. The ToOpt of a low-drag profile in turbulent flow is studied using the modified LSKE model. It is tested by comparing velocity profiles that the model can reflect the influence of solid on turbulence at Reynolds numbers as high as one million. A wall-distance computation method that can recognize the solid modeled by Darcy's source term is developed and integrated into the modified SST turbulence model. The ToOpt of a rotor-like geometry, which is of great importance in aerodynamic design, is conducted using the modified SST model, proving the model's ability to handle the ToOpt of turbomachinery.

**Key Words:** Aerodynamic design, Topology optimization, Turbulence model, Discrete adjoint

---

[1] Graduate student. Email: wcy22@mails.tsinghua.edu.cn
[2] Corresponding author, associate professor. Email: zhangyufei@tsinghua.edu.cn

# 1. Introduction

In general, aerodynamic design can be decomposed into three phases [1]: conceptual design, preliminary design, and detail design. Numerical optimization methods are widely used in the latter two phases. However, in the conceptual design phase, without the help of numerical optimization, engineers often must rely on their experience and instinct to give the initial configuration from scratch. This approach can bring many limitations to the design since the initial configuration given by the conceptual design greatly influences the overall performance of the product. Hopefully, the development of flow topology optimization (ToOpt) might change the current situation. ToOpt does not require an a priori initial condition (but any initial condition can be imposed if needed) and has a high degree of freedom that even allows a change in the connectivity of the configuration. The attributes of ToOpt make it a suitable numerical optimization method in the conceptual design phase, helping engineers search in a vast design space.

ToOpt based on Darcy's source term is a widely used method for conducting flow topology optimization. It was first introduced by Borrvall et al. [2] to Stokes flow. It uses porous media whose permeability is very small to simulate solid. The effect of the porous media on the fluid flow is characterized by a source term called Darcy's source term. The direction of Darcy's source term is always opposite to the local velocity direction, representing the resistance exerted by the porous media on the fluid. The magnitude of the source term is set to a very large value ($\kappa_{max}$) in solid regions to compel the velocity to zero, thus modeling the solid. When the CFD method is used to obtain the flow field and the value of the objective function, the distribution of Darcy's source term can be stored on the same grid throughout the optimization process, hence avoiding the need to regenerate the grid every time the solid distribution is updated. Moreover, in practice, the distribution of Darcy's source term can be treated as the design variable of ToOpt, allowing us to apply the discrete adjoint method to compute the gradient of the objective function. The above two features make ToOpt based on Darcy's source term convenient to use.

Darcy's source term was introduced to the N–S equation by Gersborg et al. [3]. The work in [4] derived the relationship between $\kappa_{max}$ needed to impede the fluid flow and the freestream velocity, characteristic length and viscosity by dimensional analysis at low and medium Reynolds numbers (laminar flow). Based on the theoretical works given by [3][4][5] and others, research on the application of ToOpt to laminar flow continued to progress. The ToOpt of a wide variety of channel flows was studied by minimizing energy dissipation as the objective function [5][6][7][8]. Some applications are related to aerodynamic design, including the ToOpt on the low-drag profile in laminar flow [9] and the ToOpt of a rotor abstracted from a centrifugal compressor at Re = 500 [10].

Recent studies have extended the scope of ToOpt to turbulent flow. A modification term similar to Darcy's source term that compels eddy viscosity to zero in solid regions was first introduced by Papoutsis-Kiachagias et al. [11] to the SA turbulence model. Gil Ho Yoon [12] and Dilgen [13] also focused on the SA turbulence model and proposed two wall-distance computation methods based on the Eiknoal equation that considers the solid modeled by Darcy's source term. Gil Ho Yoon used the modified SA model to study the ToOpt of a converging channel at $Re = 3000$, and Dilgen applied it to the ToOpt of a U-bend at $Re = 5000$. Modification terms similar to Darcy's source term were also introduced to the Wilcox $k-\omega$ model [13] and the standard $k-\epsilon$ model [14]. Both models were applied to the ToOpt of channel flows at Reynolds numbers of approximately 3000, which is quite low. To the best of the authors' knowledge, the modification related to Darcy's source term has only been added to a few of the classic turbulence models (SA, Wilcox $k-\omega$ and $k-\epsilon$) in the current literature. Some models that are widely used in industrial applications (such as the shear-stress transport (SST) model) or some popular variants of the classic models (such as the $k-\overline{v^2}-\omega$ model and Launder – Sharma $k-\epsilon$ model) still lack related modifications. On the other hand, the ability of the modified models to reflect the influence of Darcy's source term on turbulence is rarely tested. Moreover, to the best of the authors' knowledge, the relationship between the $\kappa_{max}$ needed to impede the fluid flow and the length scale, freestream velocity, and fluid viscosity at high Reynolds numbers remains unclear. The problems listed above are obstacles that need to be overcome before ToOpt exhibits its full power in aerodynamic design, where the flow encountered in practice is almost always turbulent and the Reynolds number is often high.

In this work, we focus on solving the problems of ToOpt at high Reynolds numbers in order to promote its application in the field of aerodynamic design. The structure of this paper is as follows: In Section 2, the mathematical formulation of ToOpt based on Darcy's source term is introduced, followed by the derivation of the relationship $\kappa_{max}(U, L, \nu)$ at high Reynolds numbers and the strategy for setting $\kappa_{max}$ proposed by this work. In Section 3, the modified turbulence models developed in this paper that consider the effect of Darcy's source term are introduced. In Section 4, a concise description of the discrete adjoint method used in this study to calculate the gradient of the objective function is provided. The ToOpt examples calculated using the modified turbulence models are given in Section 5.

## 2. ToOpt based on Darcy's source term

In this section, the mathematical formulation of ToOpt based on Darcy's source term is demonstrated. Then, a strategy for setting the magnitude of Darcy's source term ($\kappa_{max}$) at high Reynolds numbers is proposed based on the relationship between $\kappa_{max}$ and $U, L, \nu$. Finally, the objective functions used in this article are introduced.

### 2.1. Mathematical formulation

ToOpt based on Darcy's source term minimizes (or maximizes) the objective function defined by the user under certain constraints by searching for an optimized solid distribution. The governing equations of steady, incompressible fluid flow in ToOpt are written as:

$$\mathbf{u} \cdot \nabla \mathbf{u} = -\frac{1}{\rho}\nabla p + \nu \nabla^2 \mathbf{u} - \kappa(\alpha)\mathbf{u} \qquad (2.1)$$

$$\nabla \cdot \mathbf{u} = 0 \qquad (2.2)$$

where $\mathbf{u}, p$ are instantaneous velocity and pressure respectively. For ToOpt of turbulent flows, we use RANS equations as the governing equations:

$$\mathbf{u} \cdot \nabla \mathbf{u} = -\frac{1}{\rho}\nabla p + \nu\nabla^2 \mathbf{u} - \kappa(\alpha)\mathbf{u} + \nabla \cdot \mathbf{\tau_R} \tag{2.3}$$

$$\nabla \cdot \mathbf{u} = 0 \tag{2.4}$$

where $\mathbf{u}, p$ are averaged velocity and pressure respectively, $\tau_R$ is the Reynolds stress derived from the averaging of the nonlinear convection term in Eq. (2.1). Since $\tau_R$ makes Eq. (2.3) and Eq. (2.4) unclosed, additional equations should be added. In this study, Boussinesq hypothesis and turbulence model are introduced to close the system:

$$\mathbf{\tau_R} = 2\nu_T \cdot \frac{1}{2}(\nabla u + u\nabla) - \frac{2}{3}kI \tag{2.5}$$

$$T(\nu_T, u, p, \alpha, k, \cdots) = 0 \tag{2.6}$$

where $\nu_T$ is the eddy viscosity, $I$ is the second order identity tensor, $T(\cdot)$ is the abstract representation of turbulence model and $k, \cdots$ stands for turbulence variables such as turbulent kinetic energy etc. Combining Eq. (2.3) to Eq. (2.6), the averaged flow field can be solved.

$-\kappa(\alpha)\mathbf{u}$ in momentum equation (Eq. (2.1), Eq. (2.3)) is called Darcy's source term, and $\alpha$ is a scalar field between 0 and 1. $\kappa(\alpha)$ satisfies:

$$\kappa(1) = \kappa_{max} \gg 1\ s^{-1}, \kappa(0) = 0 \tag{2.7}$$

Eq. (2.7) implies that in the region where $\alpha = 1$, $\kappa(\alpha)$ is very large and can compel the velocity to zero through Eq. (2.1), indicating a solid region. Physically speaking, the region where $\kappa(\alpha)$ is very large ($\kappa(\alpha) = \kappa_{max}$) can be considered a region filled with porous media with low permeability. In this study, the region where $\kappa = \kappa_{max}$ and $|\mathbf{u}|_{max} < 0.03 U_\infty$ is considered to be solid. $|\mathbf{u}|_{max}$ often emerges in the region that is directly hit by the freestream in right angle. For other "solid" area, which usually makes up the vast majority of the total "solid" region, the velocity magnitude is significantly smaller than $|u|_{max}$. In the region where $\alpha = 0$, $\kappa(\alpha) = 0$ and Eq. (2.1) regresses to the standard governing equation of fluid flow, indicating a fluid region. Consequently, the distribution of $\alpha$ can represent the distribution of solids in ToOpt. In this study, $\kappa(\alpha)$ takes the form first introduced by Borrvall et al. [2]:

$$\kappa(\alpha) = \kappa_{max} \frac{q\alpha}{1 + q - \alpha}, q > 0 \qquad (2.8)$$

$q$ controls the shape of $\kappa(\alpha)$, as shown in Figure 2.1. Borrvall et al. suggest setting a smaller $q$ (~ 0.01) in the initial phase of ToOpt to avoid local minima and increasing it (~ 0.1) as the optimization proceeds to obtain values of 0 or 1s to $\alpha$ [3].

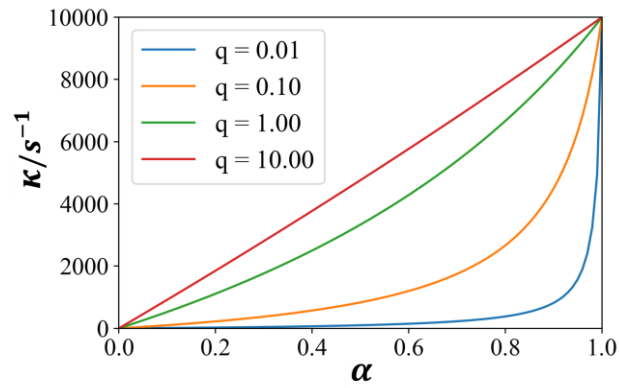

Figure 2.1 $\kappa(\alpha)$'s variation with respect to $\alpha$ for different values of $q$ when $\kappa_{max} = 1.0 \times 10^4$.

Setting Eq. (2.1) as a constraint, the mathematical formulation of ToOpt of turbulent flow based on Darcy's source term can be written as:

$$\min_{\alpha} f(\alpha, \mathbf{u}, p, \nu_T)$$
$$\text{s.t. } \mathbf{u} \cdot \nabla \mathbf{u} = -\frac{1}{\rho}\nabla p + \nu \nabla^2 \mathbf{u} + \nabla \cdot \boldsymbol{\tau}_R$$
$$\boldsymbol{\tau}_R = 2\nu_T \cdot \frac{1}{2}(\nabla \mathbf{u} + \mathbf{u}\nabla)$$
$$T(\nu_T, \mathbf{u}, p, \alpha, k, \cdots)$$
$$g(\alpha, \mathbf{u}, p, \nu_T)$$

where $f(\alpha, \mathbf{u}, p, \nu_T)$ is the objective function (functional), and $g(\alpha, \mathbf{u}, p, \nu_T)$ is the design constraint. Since field $\alpha$ is used to represent the solid distribution in ToOpt, an optimized solid distribution is found once the above optimization problem is solved for $\alpha$. In practice, the flow field is calculated by CFD, and then the objective function is evaluated. In this case, the computational domain is discretized into $n$ cells, and the functions such as $\alpha, \mathbf{u}, p$ are represented by vectors with hat ($\hat{\boldsymbol{\alpha}}, \hat{\boldsymbol{u}},$ and $\hat{\boldsymbol{p}}$) that store the respective function value on each cell. The discrete version (the discrete equations that are solved in the CFD code) of Eq. (2.3) to Eq. (2.6) can be abbreviated as:

$$R(\hat{\boldsymbol{\alpha}}, \hat{\boldsymbol{w}}) = 0 \tag{2.9}$$

$$\hat{\boldsymbol{w}} = [\hat{\boldsymbol{u}}, \hat{\boldsymbol{p}}, \hat{\boldsymbol{v}}_T, \hat{\boldsymbol{k}}, \cdots] \tag{2.10}$$

Based on Eq. (2.9) and Eq. (2.10), the discrete expression for ToOpt based on Darcy's source term can be written as:

$$\min_{\hat{\boldsymbol{\alpha}} \in S_d} \hat{f}(\hat{\boldsymbol{\alpha}}, \hat{\boldsymbol{w}})$$
$$\text{s.t. } R(\hat{\boldsymbol{\alpha}}, \hat{\boldsymbol{w}}) = 0$$
$$\hat{g}(\hat{\boldsymbol{\alpha}}, \hat{\boldsymbol{w}}) \geq 0$$

The process of solving the discrete ToOpt problem is illustrated in Figure 2.2. The functional $f$ and $g$ that generate Figure 2.2 can be found in Eq. (2.22) and Eq. (2.24). Any vector whose elements all fall between 0 and 1 can be chosen as the initial $\hat{\boldsymbol{\alpha}}$. The gradient of the objective function and the constraint are obtained by the discrete adjoint method (which will be discussed in Section 4) during the search of the optimized distribution.

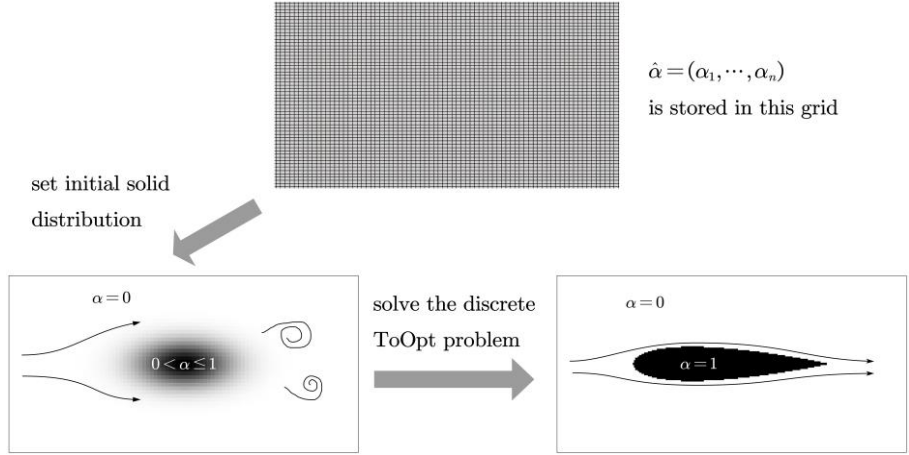

Figure 2.2   The solution process of discrete ToOpt problem.

## 2.2. Strategy for setting $\kappa_{\max}$ at high Reynolds numbers

The higher $\kappa_{\max}$ is, the more impermeable the solid modeled by Darcy's source term. However, a large $\kappa_{\max}$ induces severe stiffness to Eq. (2.1) (if the source term is treated explicitly) and to the adjoint equation that will be introduced later (Eq. (4.3)), which is an undesirable feature for numerical solution [2]. Therefore, we should seek a $\kappa_{\max}$ that is as small as possible while modeling the solid with acceptable fidelity. Ref. [4][15] suggests that for relatively low Reynolds number flow ($\text{Re} \sim 100$), $\kappa_{\max}$ should be chosen to ensure that Darcy's source term significantly outweighs the viscous force:

$$\text{Da} = \frac{\nu}{\kappa_{max} L^2} \approx 10^{-6} \sim 10^{-5} \tag{2.11}$$

Da is the ratio of the magnitude of the viscous force and Darcy's source term. In the following paragraphs, the relationship between $\kappa_{\max}$ needed to model the solid with high fidelity (If the maximum velocity in the region where $\kappa(\alpha) = \kappa_{max}$ is smaller than $\epsilon U, \epsilon \ll 1$, the solid is considered to be modeled with high fidelity. In this study, $\epsilon = 0.03$ is considered to be accurate enough because of possible direct hit of freestream on the "solid" in some optimization examples) and $U, L, \nu$ is studied ($\kappa_{max}(U, L, \nu)$). Then, a strategy for setting $\kappa_{\max}$ is proposed.

The derivation of $\kappa_{max}(U, L, \nu)$ needs two hypotheses:

**Hypothesis A:** When Darcy's source term sufficiently impedes the fluid motion in the solid region ($|\mathbf{u}| < \epsilon U, \epsilon \ll 1$, $U$ is the freestream velocity), $\mathbf{u}, p$ approximately satisfies the following relations (Darcy's equation and the continuity equation) in the solid region:

$$-\nabla p - \kappa_{\max}\mathbf{u} = 0, \nabla \cdot \mathbf{u} = 0 \qquad (2.12)$$

**Hypothesis B:** When Darcy's source term sufficiently impedes the fluid motion in the solid region, the pressure is continuous across the solid–fluid boundary.

The above hypothesis can be tested in the following case. Darcy's source term is used to model an airfoil. The Reynolds number based on the chord length is $1.0 \times 10^5$(consider a fully turbulent flow induced by trips at the leading edge of the airfoil), and $|\mathbf{u}| < 0.03U$ is satisfied in the airfoil modeled by Darcy's source term (by definition, the solid is modeled with high fidelity in this case). For Hypothesis A, the continuity equation holds since the flow field is obtained by solving Eq. (2.1). Figure 2.3 shows that the $x$ component of the left-hand side of Darcy's equation in Eq. (2.12) is approximately zero in the solid region. The distribution of the $y$ component is similar. Therefore, Hypothesis A is tested to be reasonable. Figure 2.4 shows that the pressure is continuous across the solid–fluid boundary, justifying Hypothesis B.

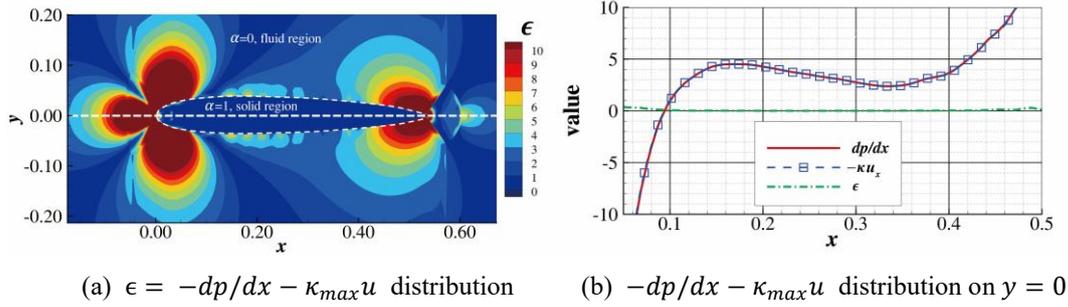

(a) $\epsilon = -dp/dx - \kappa_{max} u$ distribution  (b) $-dp/dx - \kappa_{max} u$ distribution on $y = 0$

Figure 2.3 Distribution of the left-hand side of the $x$ component of Darcy's equation in an airfoil case at $Re = 1.0 \times 10^5$.

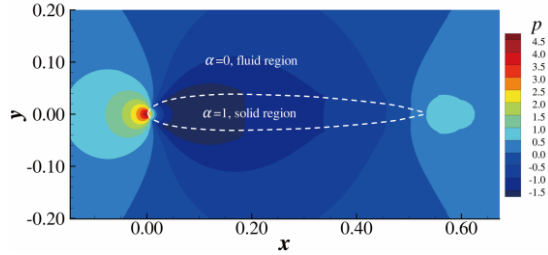

Figure 2.4 Pressure distribution of the domain.

Using hypothesis A and B, it can be derived that $p$ approximately satisfies the following boundary value problem:

$$\nabla^2 p = 0, \quad p|_{\partial\Omega} = p_{\text{airfoil}} \qquad (2.13)$$

where $\partial\Omega$ is the boundary of the solid region, and $p_{\text{airfoil}}$ is the pressure distribution on the surface of the airfoil given by solving the flow field outside of the airfoil using N–S equations with a solid wall boundary condition (without using Darcy's source term). At high Reynolds number where the flow has no separation (i.e., for airfoil, the angle of attack should be small enough to make this assumption holds), the viscous effect only dominates in the boundary layer and the boundary layer is very thin. Though the details of pressure distribution might be influenced by the displacement thickness, the outline and the extrema of $p_{airfoil}$ distribution is mainly determined by inviscid effect. Since we are just estimating the approximate value of $\kappa_{max}$, the viscous effect on pressure distribution is dropped here. Consequently, $p_{\text{airfoil}}$ is only related to the freestream velocity $U$, the length scale of the airfoil $L$, and the geometry of the airfoil $\xi$. The boundary value problem Eq. (2.13) suggests that the value of $p$ in the solid region $\Omega$ is only related to the shape of $\Omega$ and the information on the boundary ($p_{\text{airfoil}}$). Consequently, $|\nabla p|_{\max}$ is only related to $U, L,$ and $\xi$. Using dimensional analysis and the result of the above discussion, $|\nabla p|_{\max}$ can be written as:

$$|\nabla p|_{\max} = f(\xi)\frac{U^2}{L} \qquad (2.14)$$

Hypothesis A indicates that the maximum velocity magnitude in the solid region satisfies:

$$|\mathbf{u}|_{\max} = \frac{1}{\kappa_{\max}}|\nabla p|_{\max} \qquad (2.15)$$

Insert Eq. (2.14) into Eq. (2.15):

$$|\mathbf{u}|_{\max} = \frac{1}{\kappa_{\max}}f(\xi)\frac{U^2}{L} \qquad (2.16)$$

To ensure that $|\mathbf{u}| < \epsilon U$, $\kappa_{\max}$ should satisfy:

$$\frac{1}{\kappa_{\max}}f(\xi)\frac{U^2}{L} < \epsilon U \Leftrightarrow \kappa_{\max} > f(\xi)\frac{U}{\epsilon L} \Leftrightarrow \frac{\kappa_{max}L}{U} > \frac{f(\xi)}{\epsilon} \qquad (2.17)$$

Eq. (2.17) suggests that the minimal $\kappa_{max}$ needed to resist the flow is proportional to $U$. Based on Eq. (2.17), the following strategy for setting $\kappa_{max}$ is proposed:

1. Set $\kappa_{max}$ based on Eq. (2.11) as an initial guess.

2. If the initial guess is insufficient to impede the fluid motion in the solid, then $\kappa_{max}$ is increased until the desired accuracy is obtained (the velocity in the region where $\kappa(\alpha) = \kappa_{max}$ is smaller than $\epsilon U$, where $\epsilon \sim 0.01$ and is determined by the user) and the numerical solution process of the governing equations and the adjoint equations is stable.

3. Suppose we know that $\kappa_{max} = \kappa_0$ is enough for the ToOpt of one geometry at a low Reynolds number $\mathrm{Re} = \mathrm{Re}_0$, and we want to do ToOpt on the same geometry at a high Reynolds number $\mathrm{Re}^* = \beta \mathrm{Re}_0$ ($\beta > 1$). Instead of increasing the flow speed, we can keep $U$ and $\kappa_{max}$ unchanged and adjust the fluid viscosity $\nu^* = (1/\beta)\nu$ ($\mathrm{Re}^* = UL/\nu^* = \beta UL/\nu = \beta \mathrm{Re}_0$). To express the above criteria in dimensionless form, if we keep the dimensionless number $\kappa_{max} L/U$ unchanged (see Eq. (2.17)), the accuracy of solid modeling would not change regardless of the variation of Reynolds number. This strategy avoids the larger $\kappa_{max}$ required by increasing $U$, which will not change the stiffness of the N–S equation. The final solution will be the same in the dimensionless form because of the similarity law of flow.

## 2.3. Objective functions and constraints

The following objective functions and constraints are used in this paper:

1. The total pressure loss, where $\Omega$ is the computational domain, and **n** is the outer normal of the domain boundary $\partial\Omega$, is given as [6][16]:

$$\Delta P_L = -\int_{\partial\Omega}\left(p+\frac{\rho|\mathbf{u}|^2}{2}\right)(\mathbf{u}\cdot\mathbf{n})dS \qquad (2.18)$$

The total pressure loss can measure the energy dissipation in the computational domain. However, when the grid resolution near the boundary of the computational domain is coarse, the accuracy of Eq. (2.18) might be compromised.

2. The integration of the total dissipation rate [3][13] is as follows:

$$\Phi = \int_\Omega (\phi_{visc} + \phi_{Darcy})d\Omega \qquad (2.19)$$

where $\phi_{visc}$ and $\phi_{Darcy}$ are defined as:

$$\phi_{\text{visc}} = \int_\Omega \frac{1}{2}(\nu+\nu_T)(\nabla\mathbf{u}+\nabla\mathbf{u}^T):(\nabla\mathbf{u}+\nabla\mathbf{u}^T)d\Omega \qquad (2.20)$$

$$\phi_{\text{Darcy}} = \int_\Omega \kappa(\alpha)|\mathbf{u}|^2 d\Omega \qquad (2.21)$$

where $\phi_{visc}$ is the viscous dissipation rate, and $\phi_{Darcy}$ is the friction dissipation rate caused by Darcy's source term. Eq. (2.19) is an alternative to Eq. (2.18) to measure the energy dissipation in the computational domain. In practice, the separation or recirculating area far from wall contribute a lot to the unnecessary energy loss and they are the target we want to eliminate by ToOpt. Eq. (2.20) is designed to measure such unnecessary dissipation. Intermediate density ($\alpha \sim 0.5$) also causes energy loss and it is measured in Eq. (2.21).

3. The approximate drag exerted on the porous media is represented by Darcy's source term proposed by [9]:

$$D = \int_\Omega \kappa(\alpha) u_x d\Omega \tag{2.22}$$

It can be proved mathematically [9] that Eq. (2.22) tends to the following equation as the velocity in the region where $\kappa(\alpha) = \kappa_{max}$ tends to zero:

$$F_x = \left[ \int_{\partial\Omega_s} \mathbf{n} \cdot \left( -\frac{p}{\rho}\mathbf{I} + \mathbf{T} \right) d\,\partial\Omega_s \right] \cdot \mathbf{e}_x \tag{2.23}$$

where $\mathbf{T}$ is the combinition of viscous stress tensor and the Reynolds stress tensor, $\mathbf{I}$ is the second order identity tensor and $\mathbf{e}_x$ is the unit vector in $x$ direction. $\partial\Omega_s$ is the boundary of the solid region. It can be seen that Eq. (2.23) is the conventional formula used to calculate the drag exerted on a solid surface.

In some ToOpt examples of the following sections, the volume constraint of the solid is applied:

$$\eta = \frac{1}{|\Omega|} \int_\Omega \alpha d\Omega \geq \xi, \xi \in (0,1) \tag{2.24}$$

where $|\Omega|$ is the total volume of the computational domain, and $\xi$ is the lower bound of the volume fraction of the solid. In the optimization that uses drag (Eq. (2.22)) as the objective, Eq. (2.24) is used to prevent the algorithm from reducing the drag by removing the solid material. On the other hand, Eq. (2.24) can also suppress intermediate density ($\alpha \sim 0.5$) in some cases [2].

## 3. Turbulence models with modification terms related to Darcy's source term

In this paper, two modified turbulence models that can consider the influence of Darcy's source term are developed. To the best of the authors' knowledge, neither of these models has been proposed in the current literature.

### 3.1. Modified Launder – Sharma $k - \epsilon$ model

The transport equations of the modified Launder – Sharma $k - \epsilon$ (LSKE) turbulence model [17] are:

$$\frac{\partial k}{\partial t} + \frac{\partial (U_j k)}{\partial x_j} = 2\nu_T S^2 + \frac{\partial}{\partial x_j}\left(\left(\nu + \frac{\nu_T}{\sigma_k}\right)\frac{\partial k}{\partial x_j}\right) - \left(\tilde{\epsilon} \underbrace{+ D_\epsilon}\right) \underline{- c_k \kappa(\alpha) k} \quad (3.1)$$

$$\frac{\partial \tilde{\epsilon}}{\partial t} + \frac{\partial (U_j \tilde{\epsilon})}{\partial x_j} = 2C_{\epsilon_1}\frac{\tilde{\epsilon}}{k}\nu_T S^2 + \frac{\partial}{\partial x_j}\left(\left(\nu + \frac{\nu_T}{\sigma_\epsilon}\right)\frac{\partial \tilde{\epsilon}}{\partial x_j}\right) - C_{\epsilon_2} \underbrace{f_2} \frac{\tilde{\epsilon}^2}{k} + \underbrace{E_\epsilon} \quad (3.2)$$

The terms with underbraces are proposed by Launder and Sharma and are defined as:

$$D_\epsilon = 2\nu \left|\nabla\sqrt{k}\right|^2 \quad (3.3)$$

$$E_\epsilon = 2\nu\nu_T |\nabla S|^2, S = \left(2S_{ij}S_{ij}\right)^{0.5} \quad (3.4)$$

$$f_2 = 1 - 0.3\exp(-Re_t^2), Re_t = k/(\nu\tilde{\epsilon}) \quad (3.5)$$

The turbulent viscosity is calculated by:

$$\nu_T = C_\mu f_\mu k^2/\tilde{\epsilon}, C_\mu = 0.09, f_\mu = \exp[-3.4/(1 + Re_t/50)^2] \quad (3.6)$$

The standard $k - \epsilon$ model [18] necessitates the use of a wall function. The modification terms of Launder and Sharma enable the LSKE model to be integrated into the wall. Consequently, one can simply set $k, \tilde{\epsilon}$ to zero at the wall without activating the wall function when using the LSKE model.

The term underlined in the $k$ equation is proposed by the present paper and is inspired by Gil Ho Yoon's work [14]. It is quite similar to Darcy's source term in the N–S equation, and it compels $k$ to zero in the solid region (which means there is no turbulence in the solid region.). When $k$ is compelled to zero in the solid region, $\tilde{\epsilon}$ is also compelled to zero since the production term in $\tilde{\epsilon}$'s transport equation vanishes if $k = 0$:

$$P_{\tilde{\epsilon}} = 2C_{\epsilon_1}(\tilde{\epsilon}/k)\nu_T S^2 = 2C_{\epsilon_1} C_\mu f_\mu k S^2 \propto k \qquad (3.7)$$

Note that although $k$ and $\tilde{\epsilon}$ are both small in the solid region, $\nu_T$ in the solid region is still calculated by the definition in Eq. (3.6) because the numerical test suggests that $k$ tends to zero much faster than $\tilde{\epsilon}$ at the solid-fluid interface. The model's ability to describe the influence of Darcy's source term on turbulence is tested in Section 5.

## 3.2. Modified SST model

The shear-stress transport (SST) turbulence model is widely used in industry. The transport equation of the SST turbulence model [19] is listed below:

$$\frac{\partial k}{\partial t} + \frac{\partial(U_j k)}{\partial x_j} = \tilde{P}_k - \beta^* k\omega + \frac{\partial}{\partial x_j}\left[(\nu + \sigma_k \nu_T)\frac{\partial k}{\partial x_j}\right] \underline{-c_k \kappa(\alpha) k} \qquad (3.8)$$

$$\frac{\partial \omega}{\partial t} + \frac{\partial (U_j \omega)}{\partial x_j} = \alpha S^2 - \beta \omega^2 + \frac{\partial}{\partial x_j}\left[(\nu + \sigma_\omega \nu_T)\frac{\partial \omega}{\partial x_j}\right]$$
$$+ 2(1-F_1)\sigma_{w2}\frac{1}{\omega}\frac{\partial k}{\partial x_i}\frac{\partial \omega}{\partial x_i}\underline{-c_\omega \kappa(\alpha)(\omega - \omega_b)},$$
(3.9)

$\omega_b$ is defined as $800\nu/y_{min}^2$, where $y_{min}$ is the smallest grid scale. Since $\omega$ tends to infinity near the wall, the source term $-c_\omega \kappa(\alpha)(\omega - \omega_b)$ is constructed to compel $\omega$ to $\omega_b$ in solid regions, where $\omega_b$ is a very large value. The underlined terms were originally used by Dilgen et al. in the modified Wilcox $k - \omega$ turbulence model. However, different from the Wilcox $k - \omega$ model, another fix is needed in the blending function $F_1$ used by the $\omega$ equation of the SST turbulence model. By using the blending function $F_1$ in $\omega$'s transport equation, the SST model switches between the standard $k - \epsilon$ model (activated far from the wall) and the Wilcox $k - \omega$ model (activated near the wall). This zonal formulation makes the SST model robust to use and has a good resolution near the wall. $F_1$ is defined as:

$$F_1 = \tanh\left\{\left\{\min\left[\max\left(\frac{\sqrt{k}}{\beta^* \omega d}, \frac{500\nu}{d^2 \omega}\right), \frac{4\sigma_{\omega 2}k}{CD_{k\omega}d^2}\right]\right\}^4\right\} \quad (3.10)$$

It is a function of the wall-distance field $d$ ($d$ at a space point $\boldsymbol{x}$ is the distance between $\boldsymbol{x}$ and the nearest wall). $d$ field (especially near the wall) is important for the performance of the SST model since $d$ is a crucial element in the zonal formulation of the model. Traditional methods used to calculate $d$ have trouble recognizing the solid modeled by Darcy's source term, hence giving erroneous values around the solid region in ToOpt based on Darcy's source term. To address this problem, a concise method based on the p-Poisson equation and normalization proposed by [21] is developed. It contains two major steps:

1. Solve the p-Poisson equation with the penalization term proposed by this paper:

$$\nabla \cdot (|\nabla u|^{p-2} \nabla u) = -1 + \psi_{\max} \alpha u \tag{3.11}$$

   $p$ is larger than or equal to 2, and $\psi_{\max}$ is a constant.

2. Normalize $u$ to acquire $d$:

$$d = -|\nabla u|^{p-1} + \left(\frac{p}{p-1} u + |\nabla u|^p\right)^{\frac{p-1}{p}} \tag{3.12}$$

In this paper, $p$ is set to 2, and the resulting 2-Poisson equation, according to Eq. (3.11), is linear. The accuracy of the proposed method is checked in the following test case. The wall-distance field $d$ is computed by the new method in the domain shown in Figure 3.1(a). The gray area is the solid modeled by Darcy's source term. The calculated $d$ is shown in Figure 3.1(b). The result extracted from $x = -0.35, y = 0$ is plotted in Figure 3.2. The result given by the proposed method is almost identical to the analytical solution at $y = 0$. However, it deviates from the analytical solution at $x = -0.35$, where sharp corners appear in the analytical solution. Since the SST model only uses the wall-distance field to switch from the $k - \omega$ model, which is preferable within the boundary layer, to the $k - \epsilon$ model, the accuracy of the wall-distance field near the wall (around the boundary layer) is of utmost importance. The wall-distance field calculated by the proposed method overlaps the analytical solution in Figure 3.2(a) and Figure 3.2(b) near the wall, so the accuracy of the proposed method is sufficient. A coarser grid that uses 100 cells in each edge is also used to test the current method. The result in Figure 3.3 shows that the proposed scheme still performs well in this coarser grid. The validation case in section 4.2.2 further proves the accuracy of the proposed approximate wall-distance computation method.

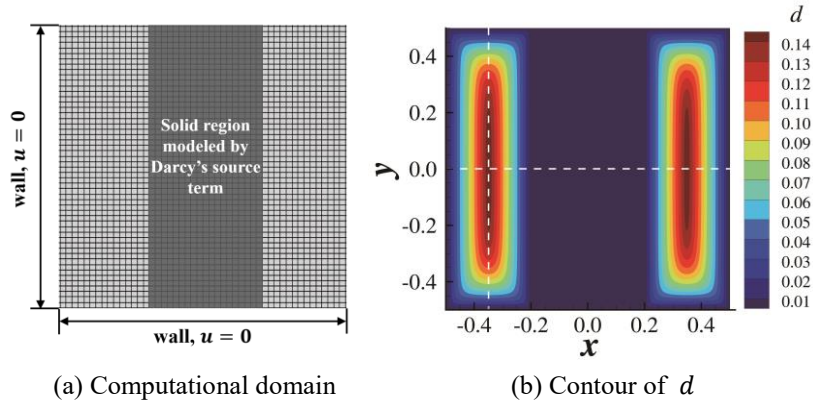

(a) Computational domain  (b) Contour of $d$

Figure 3.1 Diagram of the computational domain and the result. A $200 \times 200$ gird is used. For clarity, a coarsened grid is shown in (a).

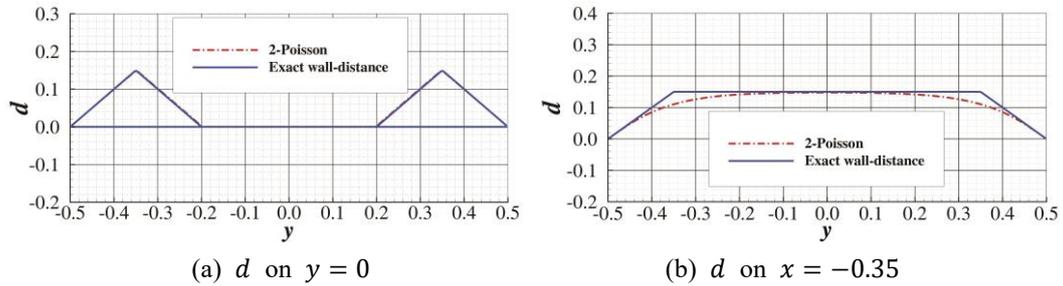

(a) $d$ on $y = 0$  (b) $d$ on $x = -0.35$

Figure 3.2 Comparison of the result computed by the proposed method and the analytical solution.

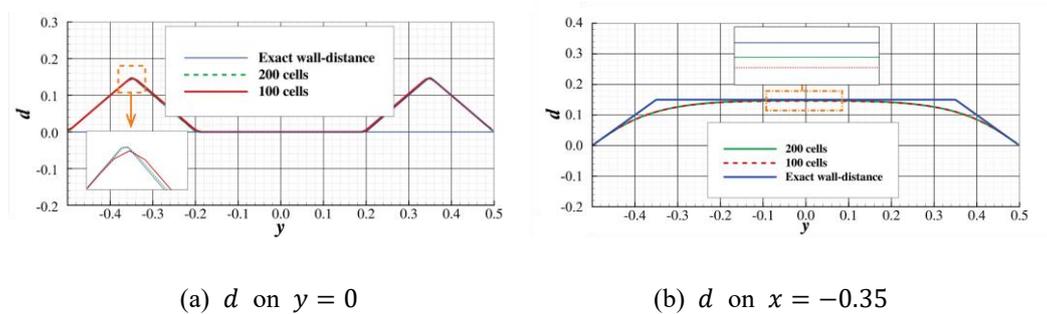

(a) $d$ on $y = 0$  (b) $d$ on $x = -0.35$

Figure 3.3 Comparison of the result given by the finer grid ($200 \times 200$) and the coarser grid ($100 \times 100$). Two results are almost identical.

# 4. Numerical method and validation

## 4.1. Numerical method for CFD and gradient computation

In this study, the open-source multidisciplinary optimization program DAFoam [22][23][24] is used to solve the CFD problem and obtain the gradient of the objective function in ToOpt. The calculated objective function value and the gradient are then fed into the optimization solver pyOptSparse [25] using SNOPT's SQP algorithm [26]. The ToOpt problem is solved by following the steps below [27][28]. The notation follows the discrete form of the ToOpt problem in Section 2.1.

1. Given the solid distribution $\hat{\boldsymbol{\alpha}}$, the state variable $\hat{\boldsymbol{w}}$ (including $\hat{\boldsymbol{u}}, \hat{\boldsymbol{p}}$ and turbulence variables) is calculated by solving the discrete form of the governing equations, including the momentum equation, the continuity equation and the turbulence model equations:

$$R(\hat{\boldsymbol{\alpha}}, \hat{\boldsymbol{w}}) = 0 \qquad (4.1)$$

Eq. (4.1) in fact represents the system of linear equations resulted from discretization. Take the momentum equation (2-D) for example, the discretized momentum equation in $x$ direction on cell $P$ can be written as:

$$\left[\frac{u_P}{2\Delta x}\left(sign(\widehat{u_e}) + sign(\widehat{u_w})\right) + \frac{v_P}{2\Delta y}\left(sign(\widehat{v_n}) + sign(\widehat{v_s})\right)\right]u_P$$

$$+ \frac{u_P}{2\Delta x}(1 - sign(\widehat{u_e}))u_E - \frac{u_P}{2\Delta x}(1 + sign(\widehat{u_w}))u_W$$

$$+ \frac{v_P}{2\Delta y}(1 - sign(\widehat{v_n}))u_N - \frac{v_P}{2\Delta y}(1 + sign(\widehat{v_s}))u_S$$

$$= -\frac{p_E - p_W}{2\Delta x} - \kappa(\alpha_P)u_P$$

(4.2)

where the variables with capital letter subscript are variables defined on cell center and the variables with lowercase letter subscript are the interpolated variables on the interface. $\Delta x, \Delta y$ are the scale of the grid cell. $u$ is the $x$ direction velocity and $v$ is $y$ direction velocity. To avoid checker-board pressure distribution, momentum interpolation might be introduced to get the $\hat{u}_i, \hat{v}_i$ [29]. The readers who are interested in the detailed and full definition of $R(\hat{\boldsymbol{\alpha}}, \hat{\boldsymbol{w}})$ are referred to the article devoted to the Discrete-Adjoint method [22].

After solving Eq. (4.1), the value of objective function $f(\hat{\boldsymbol{\alpha}}, \hat{\boldsymbol{w}})$ and constraint $g(\hat{\boldsymbol{\alpha}}, \hat{\boldsymbol{w}})$ is calculated.

2. The Jacobians $\partial R/\partial \hat{\boldsymbol{w}}, \partial R/\partial \hat{\boldsymbol{\alpha}}$, $\partial f/\partial \hat{\boldsymbol{w}}, \partial f/\partial \hat{\boldsymbol{\alpha}}$, and $\partial g/\partial \hat{\boldsymbol{w}}, \partial g/\partial \hat{\boldsymbol{\alpha}}$ are computed using automatic differentiation (AD) [30][31], which is designed to evaluate the derivative of any function specified by a computer program. The core idea of AD is that for any program, no matter how complicated it is, its output is always defined by a series of elementary operations for which the derivative is easy to compute. Due to this characteristic, chain-rule can be applied repeatedly to these operations to evaluate the derivative of the final output of the program with respect to the inputs. By using this strategy, AD can compute the derivative of any computer program accurately to machine precision.

3. The adjoint vector $\lambda_f, \lambda_g$ is calculated by solving the linear equations (adjoint equations):

$$\left(\frac{\partial R}{\partial \widehat{\boldsymbol{w}}}\right)^T \lambda_f = \left(\frac{\partial f}{\partial \widehat{\boldsymbol{w}}}\right)^T, \left(\frac{\partial R}{\partial \widehat{\boldsymbol{w}}}\right)^T \lambda_g = \left(\frac{\partial g}{\partial \widehat{\boldsymbol{w}}}\right)^T \qquad (4.3)$$

4. The gradients of the objective function and the constraint are computed:

$$\nabla f = \frac{\partial f}{\partial \widehat{\boldsymbol{\alpha}}} - \lambda_f^T \frac{\partial R}{\partial \widehat{\boldsymbol{\alpha}}}, \nabla g = \frac{\partial g}{\partial \widehat{\boldsymbol{\alpha}}} - \lambda_g^T \frac{\partial R}{\partial \widehat{\boldsymbol{\alpha}}} \qquad (4.4)$$

5. $f, g, \nabla f$ and $\nabla g$ are fed into pyOptSparse, and the SQP algorithm is used to update the solid distribution $\widehat{\boldsymbol{\alpha}}$.

Steps 1 to 5 are repeated until the convergence criteria specified for pyOptSparse is achieved. Note that the magnitude of Darcy's source term $\kappa_{max}$ always appears on the diagonal of $\partial R/\partial \widehat{\boldsymbol{w}}$, which will increase the condition number of the matrix (the 'stiffness' of the equations). Consequently, smaller $\kappa_{max}$ is more favorable for the numerical solution of Eq. (4.3).

## 4.2. Validation of the CFD solver

In this study, DAFoam's DASimpleFoam solver (which is identical to the well-known simpleFoam solver of OpenFOAM) is used to solve the CFD problem. Two cases that are similar to the optimization examples in the next section will be calculated here to validate the solver.

### 4.2.1. NACA0012, LSKE model

NASA's NACA0012 validation case [32] is calculated using the LSKE model. The LSKE model will be used in the low-drag profile optimization later in section 5.1, which is similar to this validation case. Family II grid provided by NASA [33] with $449 \times 129$ cells is used to carry out the computation. The Reynolds number based on airfoil chord length is $Re = 6.0 \times 10^6$ and the angle of attack ($AOA$) is set to $0°$.

The comparison of experimental $C_p = (p - p_\infty)/(\frac{1}{2}\rho U_\infty^2)$ distribution [34] and the computed result is shown in Figure 4.1. The result given by the current solver agrees well with the experimental data. The computed drag coefficient is compared with the tripped (corresponding to fully turbulent boundary layer) experimental data obtained by Ladson [35]. The result is shown in Table 4.1. It can be seen that the relative error between the experimental data and the current result is 2.2%, suggesting that the precision of viscous force is also acceptable.

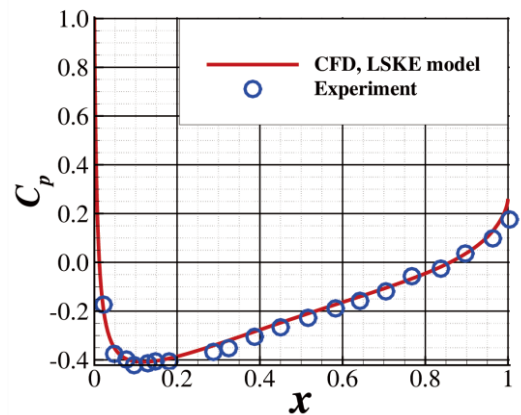

Figure 4.1 $C_p$ distribution at $AOA = 0°$

|  | Experiment[35] | Current solver |
| --- | --- | --- |
| $C_D = D/\left(1/2\rho U_\infty^2 S\right)$ | $8.14 \times 10^{-3}$ | $7.96 \times 10^{-3}$ |

Table 4.1 Drag coefficient ($C_D$) comparison

### 4.2.2. Backward facing step, $k - \omega$ SST model + 2-Poisson wall-distance approximation method

NASA's 2-D backward facing step validation case [36] is solved by DASimpleFoam using the $k - \omega$ SST model, with the wall-distance computed by the 2-Poisson equation introduced in section 3.2. Optimization example with similar geometry can be found in section 5.2. The grid generated by OpenFOAM's blockMesh utility [37] is shown in Figure 4.2. The height of the step and the channel is marked in the figure. Note that the total length of the computational domain in the streamwise direction is $130H + 50H$ (the distance between the inlet and the step + the distance between the step and the outlet) and only a part of the mesh is shown in Figure 4.2. The Reynolds number based on the height of the step is $3.6 \times 10^4$. Eq. (3.11) and Eq. (3.12) are used to compute the wall-distance field needed by $k - \omega$ SST model.

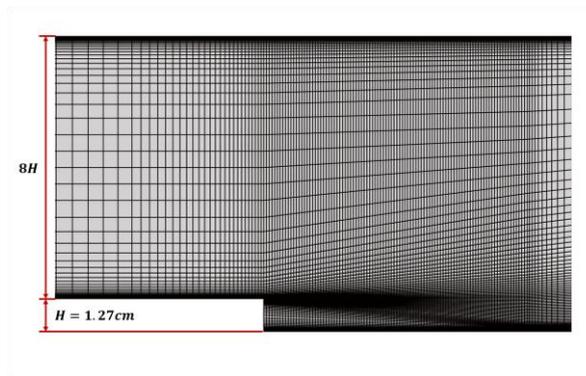

Figure 4.2 The computational grid (only part of it is shown) with 20540 cells

The computed velocity profiles at $x = H$ and $x = 4H$ downstream of the step are shown in Figure 4.3. Experimental data [39] and the result given by $k - \omega$ SST model

+ mesh-wave method [38] (the default wall-distance calculation method with high accuracy provided by OpenFOAM) are also plotted for comparison. The meaning of each line style is documented in Table 4.2 for clarity. Figure 4.3 shows that the velocity profiles obtained by mesh-wave and 2-Poisson equation are almost identical, suggesting that the accuracy of the proposed wall-distance calculation method in section 3.2 is acceptable. On the other hand, the results of CFD agree well with the experimental data. The location of reattachment point is shown in Table 4.3. The comparison suggests that the precision of reattachment point of the current solver is also satisfactory.

| Line style | Symbols | Solid line | Dashed line |
|---|---|---|---|
| **Turbulence model** | (Experimental data) | $k-\omega$ SST | $k-\omega$ SST |
| **Wall-distance** | (Experimental data) | Mesh-wave | 2-Poisson equation |

Table 4.2 The meaning of each line style in Figure 4.3

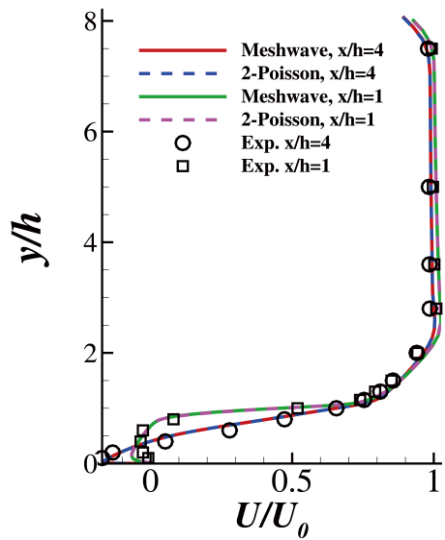

Figure 4.3 Velocity profile comparison at $x/H = 1$ and $x/H = 4$.

| | Experiment[36] | Current solver |
|---|---|---|
| $x_{reattach}/H$ | 6.26 ± 0.1 | 6.34 |

Table 4.3 The comparison of reattachment point

# 5. ToOpt examples

In this section, some ToOpt examples related to aerodynamic design are presented. Whenever the flow is assumed to be turbulent, the modified turbulence model developed in Section 3 is implemented. The ability of the modified models to depict the interaction between Darcy's source term and the turbulence is also tested in some examples.

## 5.1. Optimization of the low-drag profile

The ToOpt of the low-drag profile in the external flow was first studied by Borrvall et al. [2], assuming $Re \to 0$ (Stokes flow). It was revisited by [9], and the solution at various Reynolds numbers for laminar flow was studied. In this example, we extend the ToOpt of the low-drag profile to turbulent flow by using the modified LSKE turbulence model. In this subsection, the approximate drag (Eq. (2.22)) is chosen as the objective function:

$$D = \int_\Omega \kappa(\alpha) u_x d\Omega$$

and the volume fraction of the solid (Eq. (2.24)) is constrained to be no less than 0.03 m²:

$$\eta|\Omega| = \int_\Omega \alpha d\Omega \geq 0.03 \text{ m}^2$$

### 5.1.1. Re = 600, laminar flow

The initial solid distribution is set according to the expression below:

$$\alpha_0(x,y) = \exp\left[-\left(\frac{x^2}{a^2}+\frac{y^2}{b^2}\right)/\sigma\right] \qquad (5.1)$$

where $\sigma = 0.48$, $a = 0.25$, and $b = 0.10$. Rather than setting $\alpha = 1$, where $x^2/a^2 + y^2/b^2 < \beta\sigma$ and $\alpha = 0$ elsewhere, this initial condition can avoid vortex shedding induced by the shear layer instability while still concentrating the solid in the ellipse. The initial condition and the computational grid are visualized in Figure 5.1(a). The grid is composed of rectangular. After 100 iterations in pyOptSparse, a profile with a blunt head and sharp tail is achieved, which is quite similar to an airfoil. After the optimization process, the nondimensional approximate drag is reduced from 0.17 to 0.10. The reduction ratio is approximately 41%, showing the effectiveness of the present optimization.

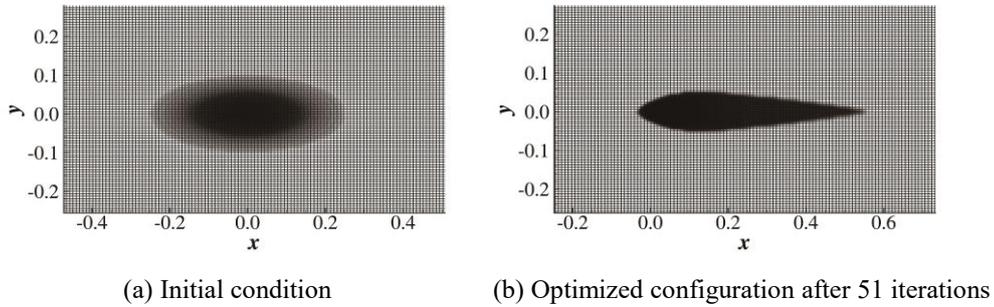

(a) Initial condition  (b) Optimized configuration after 51 iterations

Figure 5.1 Evolution of the solid distribution. Optimization is conducted on a grid composed of rectangular.

The relative thickness of the current result is compared with the results given in [9]. In the present study, the Reynolds number is calculated based on the chord length of the profile. However, in [9], the Reynolds number is based on the square root of the solid area $\sqrt{A}$. The Reynolds number based on $\sqrt{A}$ of the current result is $Re_A = 173$. The results of the optimized configuration at $Re_A = 100$ and $Re_A = 200$ are given in [9], and the corresponding relative thickness at $Re_A = 173$ is approximated using linear interpolation, as shown in Column 2 to Column 4 in Table 5.1. The interpolated result is quite similar to the current relative thickness, as shown in Column 5 in Table 5.1. Due to the lack of grid information in [9], we are not able to use the same grid here, so the comparison below is rather qualitative. The comparison shows qualitatively that for laminar flow, the higher the Reynolds number is, the thinner the optimized profile is.

|  | $Re_A = 100$ | $Re_A = 200$ | $Re_A = 173$ | Current result |
|---|---|---|---|---|
| Relative thickness | 0.189 | 0.157 | 0.167 | 0.171 |

Table 5.1 Relative thickness comparison

**5.1.2. Turbulent flow**

The low-drag profile is obtained at $Re = 3.5 \times 10^4$ and $Re = 1.0 \times 10^6$. The flow is assumed to be fully turbulent. The parameters used at different Reynolds numbers are listed in Table 5.2. $\kappa_{max}$ at $Re = 3.5 \times 10^4$ is chosen according to strategy 1 in Section 2.2 and is found to be sufficient to impede the fluid flow in the solid region ($|\mathbf{u}| \in [0, 0.03U]$ where $\kappa = \kappa_{max}$, except for a region around the leading edge directly hit by the freestream where $|\mathbf{u}| \approx 0.04U$. The scale of this region is less than 1% of the chord length so it is still included in the solid region to make the optimized configuration smooth). At $Re = 1.0 \times 10^6$, we keep $\kappa_{max}, U$ unchanged and decrease $\nu$ to acquire a higher Reynolds number according to strategy 3 in Section 2.2. This parameter setting avoids a significant increase in the equation stiffness induced by a larger $\kappa_{max}$. The intermediate solid distribution during the optimization process at $Re = 600$ is used as the initial condition.

|  | $\kappa_{max}$ | $U_\infty$ | $\nu$ |
|---|---|---|---|
| $Re = 3.5 \times 10^4$ | $1600 s^{-1}$ | 1.5 m/s | $2.25 \times 10^{-5}$ m²/s |
| $Re = 1.0 \times 10^6$ | $1600 s^{-1}$ | 1.5 m/s | $0.75 \times 10^{-6}$ m²/s |

Table 5.2 Parameters at different Reynolds numbers

The optimized configurations are presented in Figure 5.2. Both configurations have a blunt nose, a sharp tail, and a smaller thickness compared with their laminar counterparts. As shown in Figure 5.3, the nondimensional, approximate drag decreases by 56% and 30% at $Re = 3.5 \times 10^4$ and $Re = 1.0 \times 10^6$, respectively, indicating the effectiveness of the optimization process.

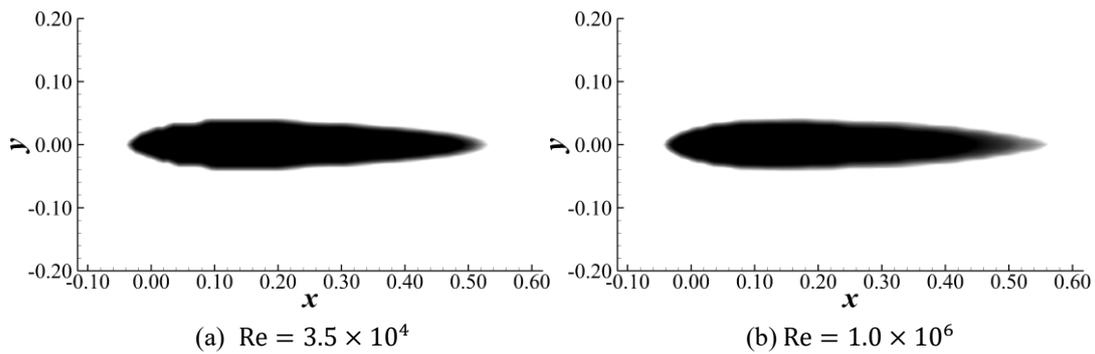

(a) $Re = 3.5 \times 10^4$  (b) $Re = 1.0 \times 10^6$

Figure 5.2 Optimized low-drag profile in turbulent flow.

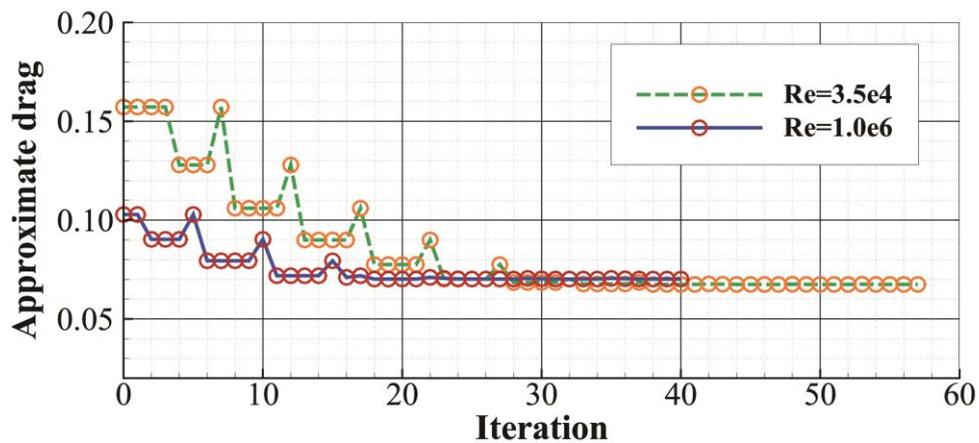

Figure 5.3 Convergence history of the nondimensional, approximate drag with respect to minor iterations.

The class-shape transformation (CST) [40] method is then applied to extract a smooth profile from the optimized configuration obtained by ToOpt, as shown in Figure 5.4. The optimized profiles at different Reynolds numbers are similar, which is partly caused by treating the boundary layer as fully turbulent in both cases. The optimized profile at $Re = 1.0 \times 10^6$ is slightly thinner than its counterpart at $Re = 3.5 \times 10^4$. Both profiles are slightly thicker than the NACA0012 airfoil, but the location of the maximum thickness is quite similar (at approximately $x = 0.3$). The resemblance between the smoothed profile and the existing widely used airfoil indicates that the current result is physically reasonable.

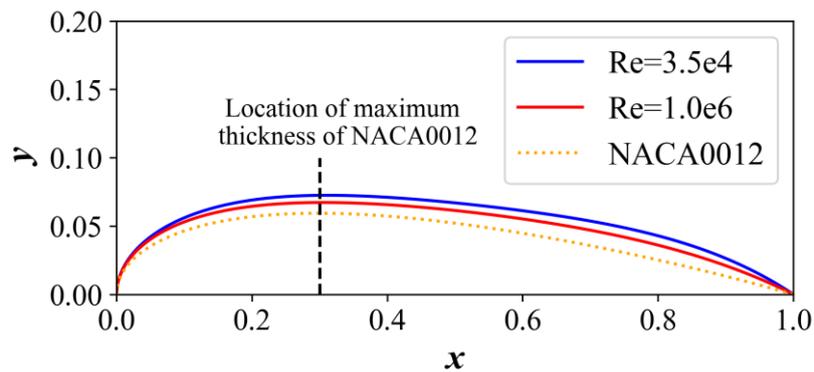

Figure 5.4 Smoothed optimized profile.

The flow field around the smoothed profile is obtained by the modified LSKE model with Darcy's source term to model the solid and the original LSKE model with a traditional solid wall boundary condition. Two kinds of body-fitted meshes are generated for solid wall computation and Darcy's source term computation. As shown in Figure 5.5, the only difference between the two meshes is that the solid domain is also covered by a grid in the mesh for Darcy's source term computation. The velocity profile and the eddy viscosity profile are plotted in Figure 5.6 and Figure 5.7. For clarity, the calculation method used for each line is documented in Table 5.3. The profiles are almost identical. This result indicates that the modified LSKE model can reflect the influence of Darcy's source term on turbulence with high fidelity even when the Reynolds number is as high as $1.0 \times 10^6$.

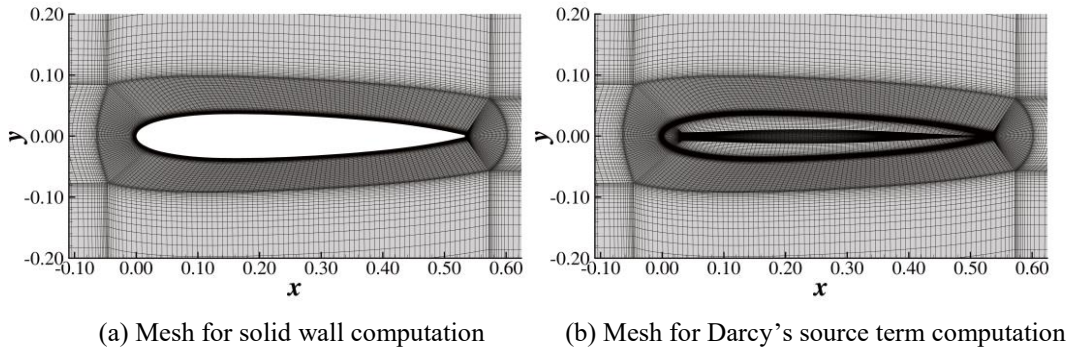

(a) Mesh for solid wall computation    (b) Mesh for Darcy's source term computation

Figure 5.5 Body-fitted mesh generated around the smoothed profile obtained at $\mathrm{Re} = 1.0 \times 10^6$. The corresponding mesh for $\mathrm{Re} = 3.5 \times 10^4$ is quite similar.

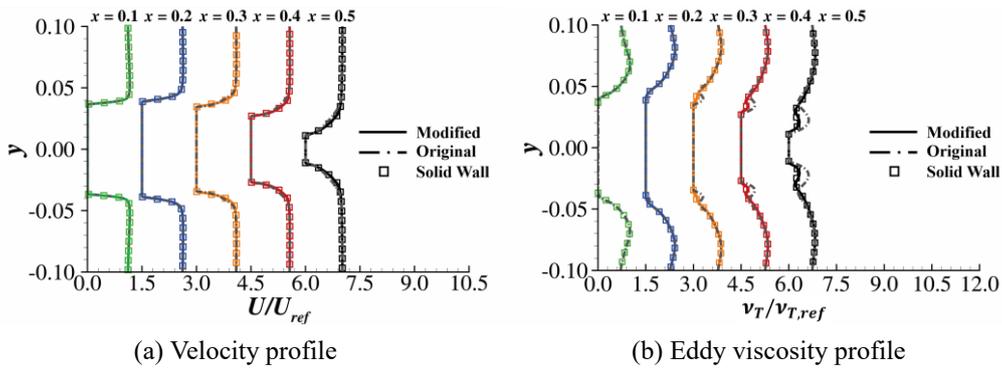

(a) Velocity profile    (b) Eddy viscosity profile

Figure 5.6 Profile comparisons at $\mathrm{Re} = 3.5 \times 10^4$; the symbols are the results given by the original LSKE model with solid wall boundary condition. For every 0.1 increase in x, the result is shifted to the right by 1.5 units.

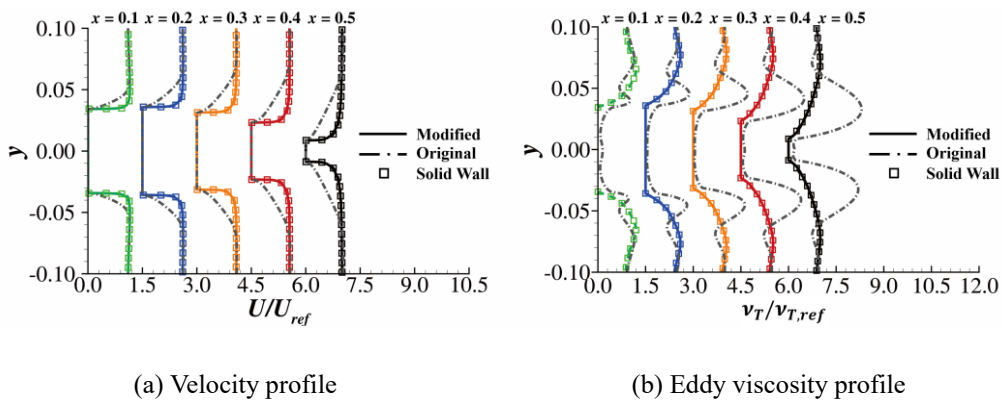

(a) Velocity profile    (b) Eddy viscosity profile

Figure 5.7 Profile comparison at $\text{Re} = 1.0 \times 10^6$. The symbols are the result given by the original LSKE model with solid wall boundary condition. For every 0.1 increase in x, the result is shifted to the right by 1.5 units.

| Line style | Square symbols | Solid line | Dashed line |
| --- | --- | --- | --- |
| **Turbulence model** | Original LSKE model | Modified LSKE model | Original LSKE model |
| **Solid representation** | Solid wall boundary | Darcy's source term | Darcy's source term |

Table 5.3 Turbulence model and the solid representation method correspond to each line style in Figure 5.7.

The necessity of the fix related to the Darcy source term in the modified LSKE model can be demonstrated as follows. Figure 5.6(b) shows that the eddy viscosity profile calculated without the fix deviates from the solid wall distribution near the wall. This deviation is much more severe at $\text{Re} = 1.0 \times 10^6$ and causes a large error in the velocity profile indicated by Figure 5.7. The deviation induced by the LSKE model without the fix can also be clearly seen in the contours of the flow variables. The shear stress profiles at $\text{Re} = 1.0 \times 10^6$ along $x = 0.3, 0.4$ are plotted in Figure 5.8. The result shows that the profile computed by the corrected LSKE model agrees well with the solid-wall solution and improves significantly compared with the original model. However, unlike the solid-wall boundary condition, the velocity is not imposed to be zero at the solid-fluid interface when Darcy's source term is used to model the solid. Consequently, the shear stress immediately adjacent to the wall is not quite accurate. Figure 5.9 compares the flow field at $\text{Re} = 1.0 \times 10^6$ computed by three methods using the contour of the field variables, which might be a more intuitive way to show the result in Figure 5.7 and Figure 5.8. The solid wall computation (baseline), Darcy's source term computation with the modified LSKE model and Darcy's source term computation with the original LSKE model. The velocity field and turbulent kinetic field calculated by the modified LKSE model agree well with the baseline. However, the boundary layer obtained by Darcy's source term with the original LSKE model is significantly thicker than the baseline, which is definitely nonphysical at $Re = 1.0 \times 10^6$. Considering the above observation, we can say that in general, the modification related to Darcy's source term in the corrected LSKE model proposed by this paper can lead to improved results.

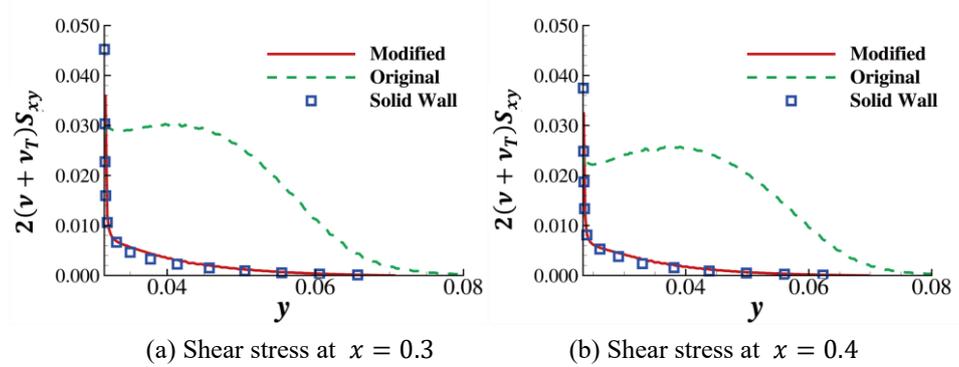

(a) Shear stress at $x = 0.3$  (b) Shear stress at $x = 0.4$

Figure 5.8 The shear stress ($\tau = 2(\nu + \nu_T)S_{xy} = (\nu + \nu_T)\left(\frac{\partial u}{\partial y} + \frac{\partial v}{\partial x}\right)$) comparison at $Re = 1.0 \times 10^6$ along the wall normal direction.

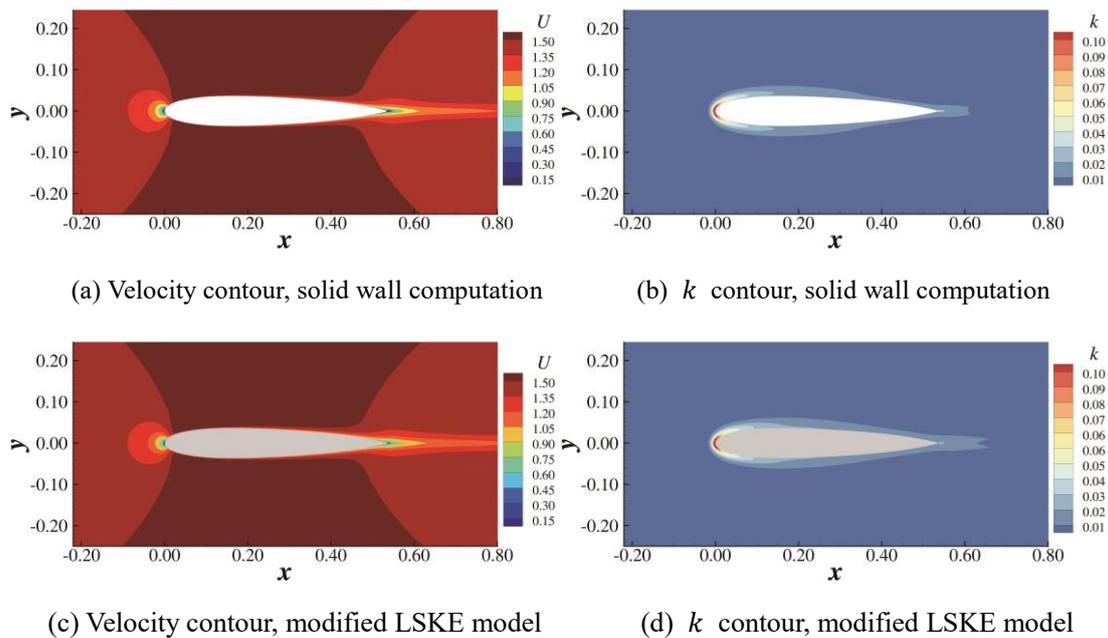

(a) Velocity contour, solid wall computation  (b) $k$ contour, solid wall computation

(c) Velocity contour, modified LSKE model  (d) $k$ contour, modified LSKE model

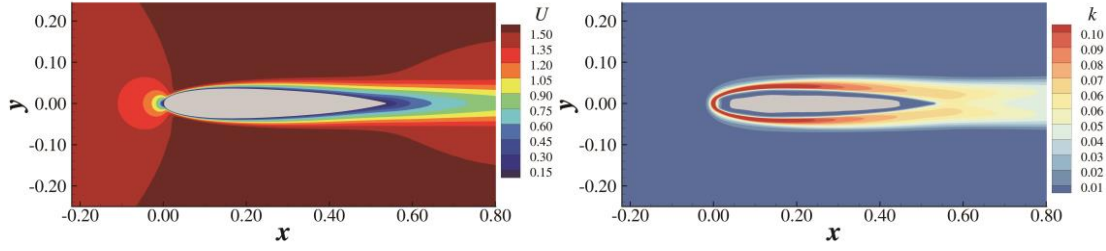

(e) Velocity contour, original LSKE model      (f) $k$ contour, original LSKE model

Figure 5.9 Comparisons of the three computations at Re $= 1.0 \times 10^6$ (right column: velocity contour; left column: turbulent kinetic energy; region where $|u| < 0.04 U_\infty$ or $k < 0.001$ m$^2$/s$^2$ is cut off ).

## 5.2. Rearward facing step

In this study, the ToOpt of the classic rearward facing step [41] is considered using the modified SST turbulence model. The objective function in this case is the total pressure loss $\Delta P_L$ (see Eq. (2.18)):

$$\Delta P_L = -\int_{\partial \Omega} \left( p + \frac{\rho |\mathbf{u}|^2}{2} \right) (\mathbf{u} \cdot \mathbf{n}) dS$$

The total pressure loss is a measure of energy dissipation. No constraints are added in this case. In other words, the algorithm will explore all configuration possible to decrease the dissipated energy in the domain. $\kappa_{max}$ is set to $1.0 \times 10^4$ based on strategy 1 in Section 2.2. The Reynolds number based on the step height is 70000. A large separation bubble is observed behind the step, as shown in Figure 5.10. The viscous dissipation rate (Eq. (2.20)) distribution is plotted in Figure 5.11.

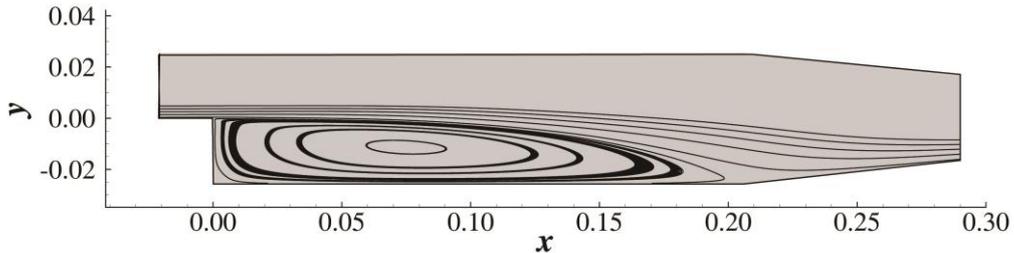

Figure 5.10 Initial condition and flow field.

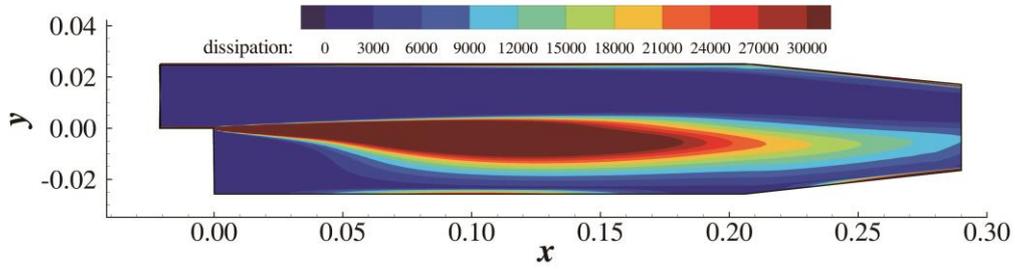

Figure 5.11 Viscous dissipation rate distribution.

The optimization converged after 19 major iterations, as shown in Figure 5.12. The reduction in the total pressure loss is shown to be 55%. In the optimized configuration, the separation bubble is eliminated by the solid distributed behind the step, as shown in Figure 5.13. Note that a geometry closer to a plane channel might result in a higher bulk velocity and thus a higher friction. So, it is reasonable to have an optimized solution shown in Figure 5.13. The viscous dissipation rate is effectively suppressed in the optimized configuration compared with Figure 5.11, as shown in Figure 5.14.

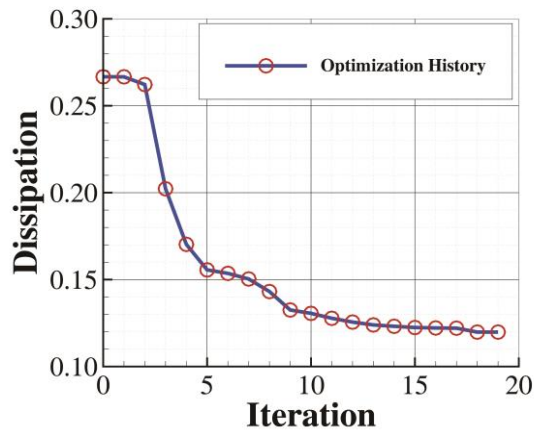

Figure 5.12 Convergence history of $\Delta P_L$.

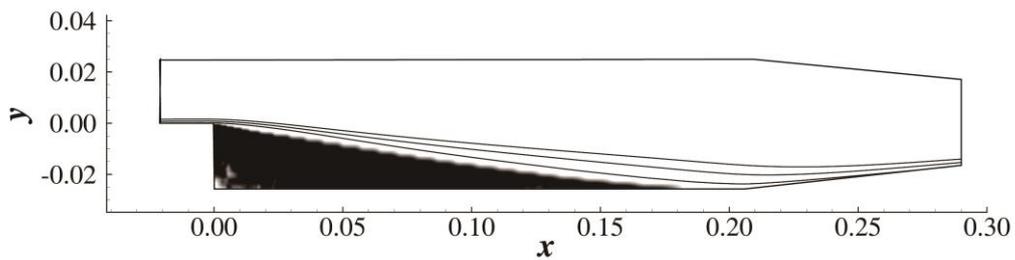

Figure 5.13 Optimized configuration.

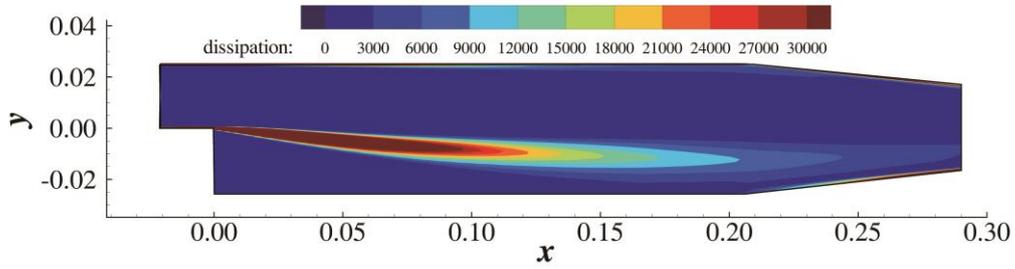

Figure 5.14 The viscous dissipation rate is largely reduced in the optimized configuration.

In Figure 5.15, a body-fitted mesh is generated around the optimized configuration. The $\Delta P_L$ of the optimized configuration is computed on this mesh and is compared with the $\Delta P_L$ of the original configuration in Table 5.4. $\Delta P_L$ is reduced by 80% in the optimized configuration, showing the effectiveness of ToOpt. Figure 5.16 shows that as $\kappa_{max}$ increases, $\Delta P_L$ computed with Darcy's source term modeling the solid approaches that computed by the body-fitted mesh. This result suggests that the SST model can effectively reflect the influence of solid material represented by Darcy's source term when $\kappa_{max}$ is large.

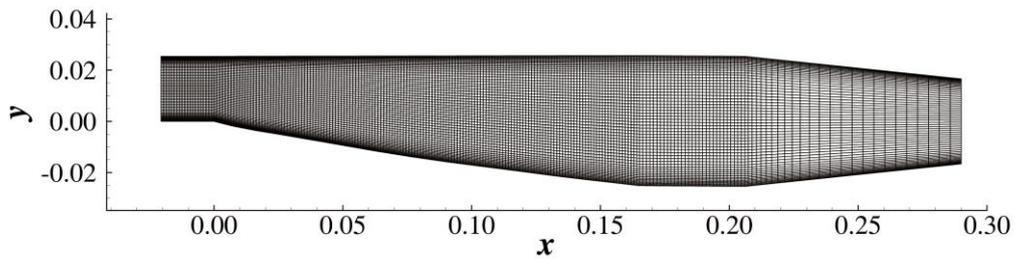

Figure 5.15 Body-fitted mesh around the optimized configuration.

|  | Original configuration | Optimized configuration |
| --- | --- | --- |
| $\Delta P_L$ | 0.2667 | 0.0532 |
| Reduction of $\Delta P_L$ | --- | 80.1% |

Table 5.4  $\Delta P_L$ comparison

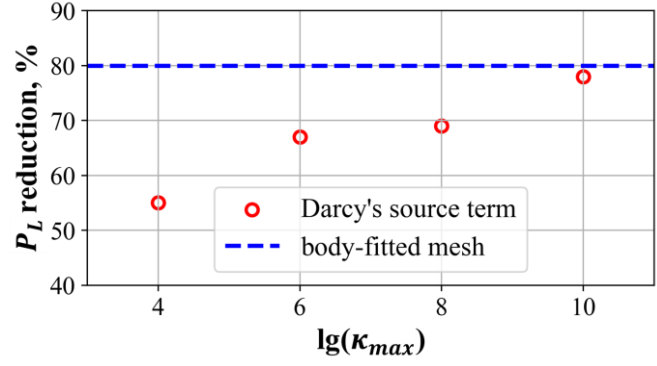

Figure 5.16 Variation in $\Delta P_L$ computed in ToOpt with respect to the log of $\kappa_{max}$.

## 5.3. Rotor-like case

Turbomachinery design is of great importance in the application of aerodynamic optimization. To show the modified SST model's ability to handle the ToOpt of turbomachinery, a geometry (Figure 5.17) [10] generalized from a centrifugal compressor (Figure 5.18) is optimized with the modified SST model. Only one inlet is included in Figure 5.17 considering the rotational symmetry of the centrifugal compressor. As shown in Figure 5.17, the whole geometry is in a noninertial frame whose rotate speed is $n = 5000$ rpm. The rotation effect is considered by adding inertial forces into the momentum equation ($\boldsymbol{\omega} = \omega \mathbf{e}_3 = 2\pi n e_3$):

$$\mathbf{u} \cdot \nabla \mathbf{u} + \underbrace{\boldsymbol{\omega} \times (2\mathbf{u} + \boldsymbol{\omega} \times \mathbf{r})} = -\frac{1}{\rho}\nabla p + \nu\nabla^2\mathbf{u} - \kappa(\alpha)\mathbf{u} + \boldsymbol{\nabla} \cdot \boldsymbol{\tau}_R \quad (5.2)$$

The integral of the dissipation rate over the whole computational domain (see Eq. (2.19)) is chosen as the objective function:

$$\Phi = \int_\Omega (\phi_{visc} + \phi_{Darcy})d\Omega$$

The volume constraint of the solid in Eq. (2.24) is activated with a lower bound of the

volume fraction of 0.8:

$$\eta = \frac{1}{|\Omega|}\int_\Omega \alpha\, d\Omega \geq 0.8$$

In the initial condition, α is set to 0.8 everywhere to fulfill the constraint. $\kappa_{max}$ is set to $1\times 10^4$ according to strategy 1 in Section 2.2.

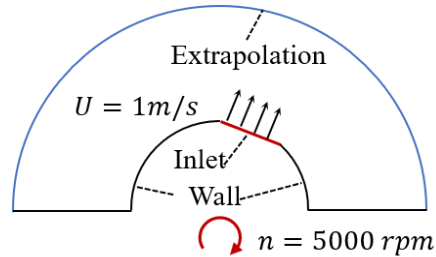

Figure 5.17 Computational domain and the boundary conditions.

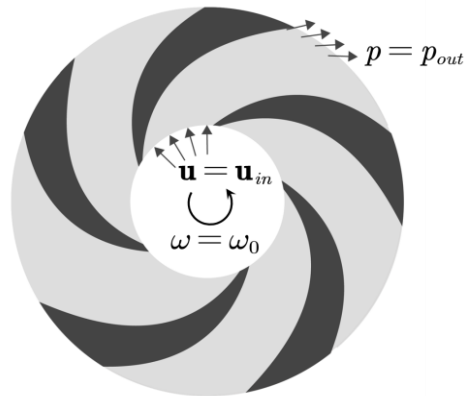

Figure 5.18 Illustrative diagram of a centrifugal compressor.

As shown in Figure 5.19, $q$ in Eq. (2.8) is set to 0.01 in the first several iterations. After the optimization converges to a solution where many cells are with intermediate α, $q$ is increased to 0.1 to make the solid–fluid boundary sharp. This trick is proposed by Borrvall et al.[2]. The optimized configuration, shown in Figure 5.20(a), is a pipe deflected toward the Coriolis force felt by the fluid particles. The pipe is copied and uniformly distributed around the inner circle to obtain a configuration resembling a centrifugal compressor, as shown in Figure 5.20(b). Figure 5.21 compares the optimized pipe and the velocity field obtained in laminar flow ($n = 500$ rpm) and the current result. The difference is fairly clear. In the laminar solution, the two sides of the pipe bend in different directions, while in the turbulent solution, both sides bend in the direction of the Coriolis force felt by the fluid particles. The velocity distribution shows that the boundary layer is much thicker in the laminar flow case than in the turbulent flow case.

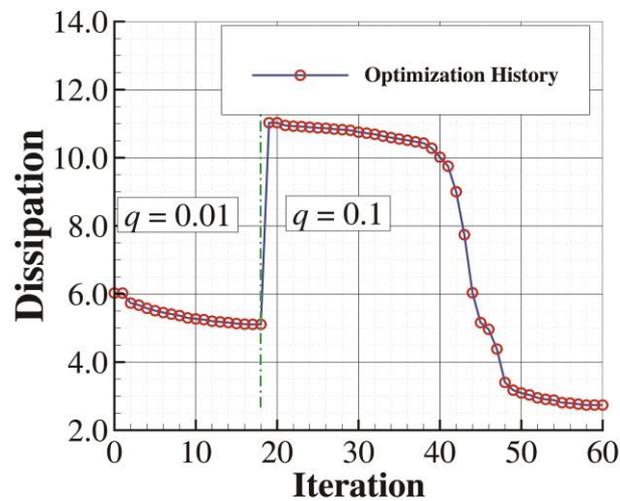

Figure 5.19 Convergence history of the energy dissipation.

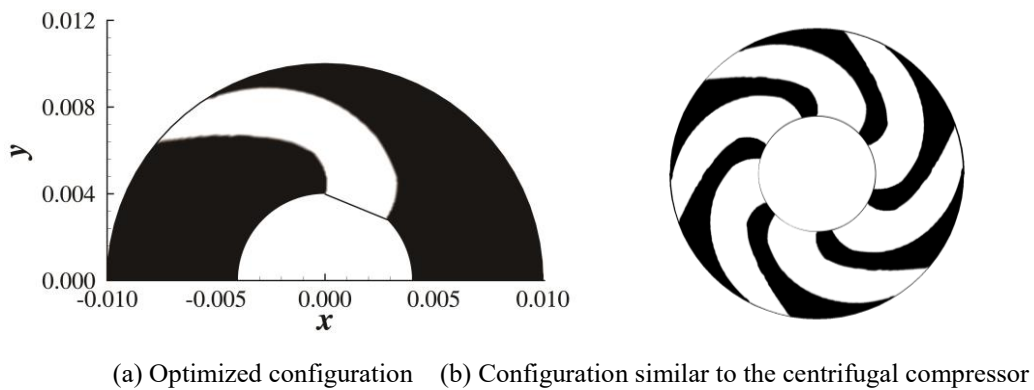

(a) Optimized configuration    (b) Configuration similar to the centrifugal compressor

Figure 5.20 Result of ToOpt.

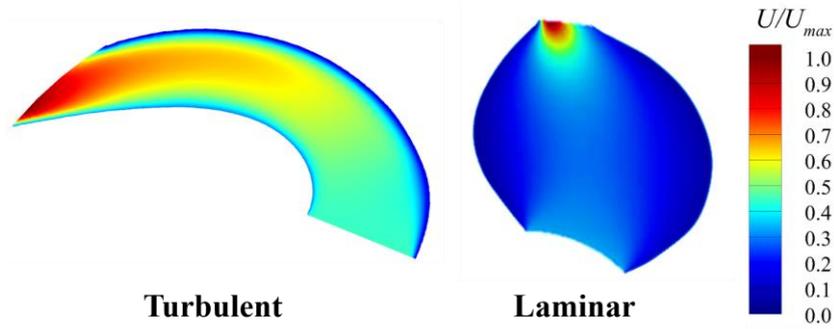

Figure 5.21 Comparison of the optimized configuration and the velocity distribution in laminar and turbulent flows. The laminar velocity distribution is extracted from [10].

The optimized configurations at other rotating speeds are also computed using ToOpt and are shown in Figure 5.22. The velocity distribution of the optimized configuration at each rotating speed is shown in Figure 5.23. All the optimized configurations are pipes deflected in the direction of the Coriolis force, and the larger the rotating speed is, the more deflected the pipe. Additionally, the maximum curvature of the pipe is shown to always occurs near the inlet.

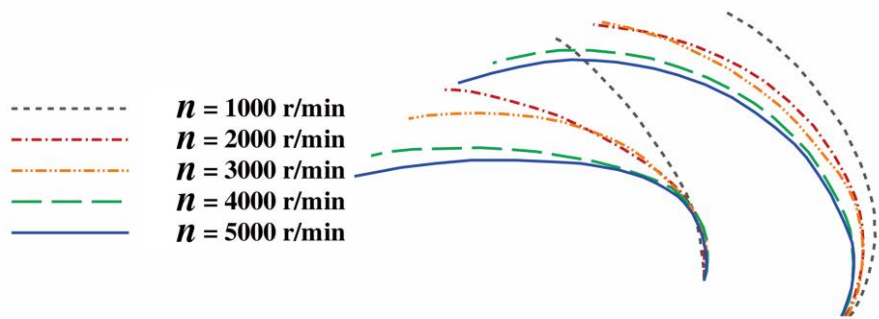

Figure 5.22 The optimized configurations are all deflected pipes.

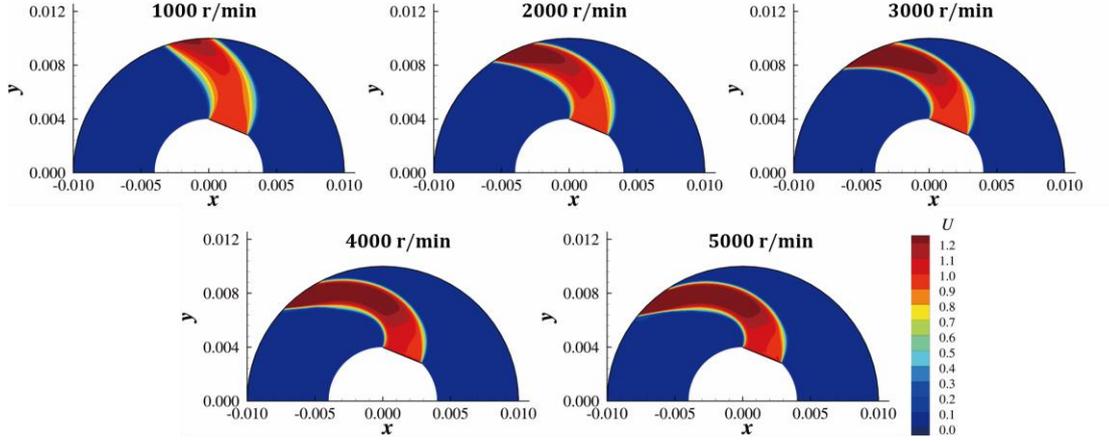

Figure 5.23 Velocity magnitude distribution of the optimized configuration at RPM = 1000, 2000, 3000, 4000 and 5000.

# 6. Conclusions

The work in this paper focuses on applying ToOpt based on Darcy's source term to the aerodynamic design under turbulent flow. The results and the problems solved can be summarized as follows:

1. The minimum $\kappa_{max}$ needed to impede the fluid flow in the solid (to make $|u| < \epsilon U$ in the solid) is proportional to the freestream velocity $U$ and is unrelated to fluid viscosity when the Reynolds number is large. Based on this relationship and previous experience [15], a strategy for setting $\kappa_{max}$ is proposed: keep the non-dimensional number $\kappa_{max} L/U$ unchanged when Reynolds number varies. The strategy is used in the examples of ToOpt and is tested to be effective.

2. The flows encountered in aerodynamic design are generally turbulent. Therefore, for ToOpt, considering the impact of Darcy's source term on turbulence is important. In this paper, a modified LSKE turbulence model is developed. The test case in subsection 5.1.2 shows that the proposed model has a satisfactory ability to depict the influence of Darcy's source term on turbulence even when the Reynolds number is as high as $1 \times 10^6$. A concise, approximate wall-distance computation method that recognizes the solid modeled by Darcy's source term is developed. This method is integrated into the modified SST model. The modified SST model can also reflect the influence of Darcy's source term on turbulence when $\kappa_{max}$ is large.

3. Many aerodynamic optimization problems are related to acquiring a configuration with the lowest drag in an external flow. ToOpt was previously mostly used to obtain the low-drag profile in laminar flow. In this study, the ToOpt of a low-drag profile is extended to turbulent flow whose Reynolds number is as high as $1 \times 10^6$, which has some significance for the application of ToOpt to aerodynamic design.

4. Optimizing turbomachinery is another important topic in aerodynamic design. In this paper, the ToOpt of a rotor-like geometry is studied using the modified SST model. The model's potential for performing the ToOpt of turbomachinery is tested. This test case shows that the larger the rotating speed is, the more deflected the optimized configuration.

The authors believe that the problems solved in this work can promote the practical application of ToOpt in aerodynamic design process where high Reynolds number turbulent flow is often encountered.

## Acknowledgments

This work was supported by the National Natural Science Foundation of China (grant nos. 11872230, 92052203 and 91952302) and the Aeronautical Science Foundation of China (grant no. 2020Z006058002).

# 7. References

[1]. Zhang Min et al. "Fluid Topology Optimization Method and Its Application in Turbomachinery" Journal of Propulsion Technology 42.11(2021):2401-2416. (in Chinese)

[2]. Borrvall, Thomas, and Joakim Petersson. "Topology optimization of fluids in Stokes flow." International journal for numerical methods in fluids 41.1 (2003): 77-107.

[3]. Gersborg-Hansen, Allan, Ole Sigmund, and Robert B. Haber. "Topology optimization of channel flow problems." Structural and multidisciplinary optimization 30.3 (2005): 181-192.

[4]. Olesen, Laurits Højgaard, Fridolin Okkels, and Henrik Bruus. "A high-level programming-language implementation of topology optimization applied to steady-state Navier–Stokes flow." International Journal for Numerical Methods in Engineering 65.7 (2006): 975-1001

[5]. Othmer, Carsten, Eugene de Villiers, and Henry Weller. "Implementation of a continuous adjoint for topology optimization of ducted flows." 18th AIAA Computational Fluid Dynamics Conference. 2007.

[6]. Pietropaoli, M., F. Montomoli, and A. Gaymann. "Three-dimensional fluid topology optimization for heat transfer." Structural and Multidisciplinary Optimization 59.3 (2019): 801-812.

[7]. Gaymann, A., F. Montomoli, and M. Pietropaoli. "Fluid topology optimization: Bio-inspired valves for aircraft engines." International Journal of Heat and Fluid Flow 79 (2019): 108455.

[8]. Othmer, Carsten. "A continuous adjoint formulation for the computation of topological and surface sensitivities of ducted flows." International journal for numerical methods in fluids 58.8 (2008): 861-877.

[9]. Kondoh, Tsuguo, Tadayoshi Matsumori, and Atsushi Kawamoto. "Drag minimization and lift maximization in laminar flows via topology optimization employing simple objective function expressions based on body force integration." Structural and Multidisciplinary Optimization 45.5 (2012): 693-701.

[10]. N Sá, L. F., et al. "Design optimization of laminar flow machine rotors based on the topological derivative concept." Structural and Multidisciplinary Optimization 56.5 (2017): 1013-1026.

[11]. Papoutsis-Kiachagias, E. M., et al. "Constrained topology optimization for laminar and turbulent flows, including heat transfer." CIRA, editor, EUROGEN, Evolutionary and Deterministic Methods for Design, Optimization and Control, Capua, Italy (2011).

[12]. Yoon, Gil Ho. "Topology optimization for turbulent flow with Spalart–Allmaras model." Computer Methods in Applied Mechanics and Engineering 303 (2016): 288-311.

[13]. Dilgen, Cetin B., et al. "Topology optimization of turbulent flows." Computer Methods in Applied Mechanics and Engineering 331 (2018): 363-393.

[14]. Yoon, Gil Ho. "Topology optimization method with finite elements based on the k-ε turbulence model." Computer Methods in Applied Mechanics and Engineering 361 (2020): 112784.

[15]. Philippi, B., and Y. Jin. "Topology optimization of turbulent fluid flow with a sensitive porosity adjoint method (spam)." arXiv preprint arXiv:1512.08445 (2015).

[16]. Zhang, Min, et al. "Aerodynamic topology optimization on tip configurations of turbine blades." Journal of Mechanical Science and Technology 35.7 (2021): 2861-2870.

[17]. Launder, Brian Edward, and Bahrat I. Sharma. "Application of the energy-dissipation model of turbulence to the calculation of flow near a spinning disc." Letters in heat and mass transfer 1.2 (1974): 131-137.

[18]. Wilcox, David C. Turbulence modeling for CFD. Vol. 2. La Canada, CA: DCW industries, 1998.

[19]. Menter, Florian R., Martin Kuntz, and Robin Langtry. "Ten years of industrial experience with the SST turbulence model." Turbulence, heat and mass transfer 4.1 (2003): 625-632.

[20]. OpenFOAM: User Guide: Wall distance calculation methods. (n.d.). Www.openfoam.com. Retrieved July 16, 2022, from https://www.openfoam.com/documentation/guides/latest/doc/guide-schemes-wall-distance.html

[21]. Belyaev, Alexander G., and Pierre-Alain Fayolle. "On variational and PDE-based distance function approximations." Computer Graphics Forum. Vol. 34. No. 8. 2015.

[22]. He, Ping, et al. "An aerodynamic design optimization framework using a discrete adjoint approach with OpenFOAM." Computers & Fluids 168 (2018): 285-303.

[23]. He, Ping, et al. "Dafoam: An open-source adjoint framework for multidisciplinary design optimization with openfoam." AIAA journal 58.3 (2020): 1304-1319.

[24]. He, Ping, et al. "An object-oriented framework for rapid discrete adjoint development using OpenFOAM." AIAA Scitech 2019 Forum. 2019.

[25]. Wu, Neil, et al. "pyOptSparse: A Python framework for large-scale constrained nonlinear optimization of sparse systems." Journal of Open Source Software 5.54 (2020): 2564.

[26]. Gill, Philip E., Walter Murray, and Michael A. Saunders. "SNOPT: An SQP algorithm for large-scale constrained optimization." SIAM review 47.1 (2005): 99-131.

[27]. Bradley, Andrew M. PDE-constrained optimization and the adjoint method. Technical Report. Stanford University. https://cs.stanford. edu/~ ambrad/adjoint_tutorial. pdf, 2013.

[28]. Duffy, Austen C. "An introduction to gradient computation by the discrete adjoint method." tech. rep. (2009).

[29]. Li, Xue-song, and Chun-wei Gu. "The momentum interpolation method based on the time-marching algorithm for all-speed flows." Journal of Computational Physics 229.20 (2010): 7806-7818.

[30]. Griewank, Andreas, and Andrea Walther. Evaluating derivatives: principles and techniques of algorithmic differentiation. Society for industrial and applied mathematics, 2008.

[31]. Sagebaum, Max, Tim Albring, and Nicolas R. Gauger. "High-performance derivative computations using codipack." ACM Transactions on Mathematical Software (TOMS) 45.4 (2019): 1-26.

[32]. "2D NACA 0012 Airfoil Validation." NASA, NASA, https://turbmodels.larc.nasa.gov/naca0012_val.html.

[33]. "Grids - NACA 0012 Airfoil for Turbulence Model Numerical Analysis." Turbmodels.larc.nasa.gov, turbmodels.larc.nasa.gov/naca0012numerics_grids.html. Accessed 15 Jan. 2023.

[34]. N. Gregory and C.L. O'Reilly. Low-speed aerodynamic characteristics of NACA 0012 aerofoil sections, including the effects of upper-surface roughness simulation hoar frost. Reports and Memoranda No. 3726, National Physics Laboratory, Teddington, UK, 1970.

[35]. C.L. Ladson. Effects of independent variation of Mach and Reynolds numbers on the low-speed aerodynamic characteristics of the NACA 0012 airfoil section. NASA Technical Memorandum NASA-TM-4074, NASA Langley Research Center; Hampton, VA, United States, 1988.

[36]. "2D Backward Facing Step." Turbmodels.larc.nasa.gov, turbmodels.larc.nasa.gov/backstep_val.html.

[37]. "OpenFOAM: User Guide: Backward Facing Step." www.openfoam.com, www.openfoam.com/documentation/guides/latest/doc/verification-validation-turbulent-backward-facing-step.html. Accessed 15 Jan. 2023.

[38]. "OpenFOAM: User Guide: Mesh-Wave Wall Distance." Www.openfoam.com, www.openfoam.com/documentation/guides/latest/doc/guide-schemes-wall-distance-meshwave.html. Accessed 15 Jan. 2023.

[39]. Nasa.gov, 2023, turbmodels.larc.nasa.gov/Backstep_validation/profiles.exp.dat. Accessed 15 Jan. 2023.

[40]. Kulfan, Brenda M. "CST universal parametric geometry representation method with applications to supersonic aircraft." Fourth International Conference on Flow Dynamics Sendai International Center Sendai, Japan. 2007

[41]. He, P. (2022, February 22). PitzDaily. DAFoam. Retrieved April 4, 2022, from https://dafoam.github.io/my_doc_tutorials_topo_pitdaily.html